\documentclass[aip,reprint,amsmath,amssymb]{revtex4-2}

\usepackage{graphicx}
\usepackage{dcolumn}
\usepackage{bm}

\usepackage[utf8]{inputenc}
\usepackage[T1]{fontenc}
\usepackage{mathptmx}

\begin{document}
\title{ F\"{o}rster resonance energy transfer in inhomogeneous and absorptive environment}
\author{L. S. Petrosyan,$^{1}$ M. N. Noginov,$^{2}$  and T. V. Shahbazyan$^{1}$} 
\affiliation{$^{1}$Department of Physics, Jackson State University, Jackson, MS 39217 USA
\\
$^{2}$Center for Materials Research, Norfolk State University, Norfolk, VA 23504, USA}


\begin{abstract} 
We present an analytical model  for  F\"{o}rster resonance energy transfer (FRET) between a donor and an acceptor placed in inhomogeneous and absorptive environment characterized by complex dielectric function, e.g., near a metal-dielectric structure. By extending the standard approach to FRET to include energy transfer (ET) channel to the environment, we show that, in the absence of plasmonic enhancement effects, the F\"{o}rster radius, which defines the characteristic distance for efficient FRET, is reduced due to a competing ET process. We demonstrate that a reduction of the F\"{o}rster radius can affect dramatically fluorescence from large ensemble of molecules whose emission kinetics is dominated by FRET-induced concentration quenching. Specifically, we perform numerical calculations for dye-doped polymer films deposited on top of metallic substrate to find that, for high dye concentrations, the emission kinetics  slows down considerably, in sharp contrast to acceleration of single-molecule fluorescence. Furthermore, the calculated effective fluorescence decay rate exhibits non-monotonic behavior with varying film thickness, consistent with the experiment, indicating a non-trivial interplay between the metal quenching and concentration quenching mechanisms. 

\end{abstract}

\maketitle



\section{Introduction}

F\"{o}rster resonance energy transfer\cite{forster-ap48,dexter-jcp53}  (FRET)  between spatially separated donor and acceptor fluorophores, e.g., dye molecules or semiconductor quantum dots, underpins diverse phenomena  such as photosynthesis, exciton transfer in molecular aggregates, or interaction between proteins.\cite{andrews-book,ag-book,lakowicz-book}  FRET spectroscopy has been widely used in numerous applications including, e.g.,  in studies of protein folding, \cite{deniz-pnas00,lipman-science03} live cell protein localization, \cite{selvin-naturesb00,sekar-jcb03} optical sensing \cite{gonzalez-bj95,medintz-naturemat03} and light harvesting. \cite{andrews-lp11} During past two decades, significant progress was made in enhancement and control of energy transfer (ET) by placing fluorophores in microcavities \cite{hopmeier-prl99,andrew-science00,finlayson-cpl01} or near metal-dielectric structures supporting plasmon resonances. \cite{leitner-cpl88,lakowicz-jf03,andrew-science04,lakowicz-jpcc07-1,lakowicz-jpcc07-2,krenn-nl08,rogach-apl08,bodreau-nl09,yang-oe09,an-oe10,lunz-nl11,zhao-jpcc12,west-jpcc12,lunz-jpcc12} While direct FRET due to dipole coupling between the fluorophores is  efficient for donor-acceptor separations below the  F\"{o}rster radius $R_{F}$ ($2\div 10$ nm for dye molecules), \cite{lakowicz-book} the plasmon-mediated ET channels  provided by metal nanoparticles, \cite{nitzan-cpl84,nitzan-jcp85,druger-jcp87,dung-pra02,stockman-njp08,pustovit-prb11,pustovit-prb13,schatz-jpcl17}  films and waveguides \cite{dung-pra02,moreno-nl10}, or charge-doped monolayer graphene, \cite{velizhanin-prb12} can significantly enhance FRET  at a much longer distances.

At the same time, quantum efficiency of a single fluorophore situated near a metal-dielectric structure is determined by the interplay between plasmonic enhancement and dissipation in metal \cite{feldmann-nl05,novotny-prl06,sandoghdar-prl06,halas-nl07} which depends  sensitively on fluorophore's proximity to the metal. \cite{nitzan-jcp81,ruppin-jcp82} In a similar way, the plasmonic enhancement of FRET between a donor and an acceptor is offset by dissipation of the excited donor's energy if its distance to the metal is sufficiently small.\cite{pustovit-prb11} In the absence of plasmon resonances, the rate of single-molecule fluorescence in inhomogeneous and absorptive environment, e.g., near a metal interface, can be strongly reduced due to emergence of ET channel to the environment.\cite{silbey-review} A similar effect is expected for FRET as well as  ET between a donor and an acceptor molecules is accompanied by ET to the environment.


On the other hand, FRET strongly affects fluorescence from large ensembles of molecules by inhibiting emission of light at high molecule concentrations. When the Stokes shift is small enough to ensure sufficient overlap between the molecules' absorption and emission bands, fluorescence of an excited molecule (donor) is accompanied by FRET to surrounding molecules (acceptors) which, for high molecule concentrations, can significantly affect the emission kinetics.\cite{huber-prb85,parkinson-prb07,krenn-nl08} For high molecule concentrations, as the average distance to nearby molecules is below the F\"{o}rster radius $R_{F}$, these processes lead to \textit{concentration quenching} of  fluorescence.\cite{noginov-nanophot21} At the same time, if molecules are placed in inhomogeneous and absorptive environment, e.g., near a metal interface, a further suppression of fluorescence would be expected due to  additional ET channel, similar to quenching of single-molecule fluorescence by the metal.
Such a behavior has indeed been reported for dye-doped polymer films deposited on top of metal substrate for low molecule concentration  as the ensemble fluorescence exhibited  faster  emission kinetics characterized by  higher effective decay rate.\cite{noginov-josa19}

However, for high molecule concentrations, the \textit{opposite} behavior has been observed  as the emission kinetics for a dye-doped polymer film on metal substrate \textit{slowed down} as compared to the same system placed on non-absorptive glass substrate.\cite{noginov-josa21} Furthermore, with varying film thickness, i.e., with changing  molecules' average distance to the metal, the effective fluorescence decay rate exhibited \textit{non-monotonic} behavior, which was  in sharp contrast to a substantially higher and nearly constant rate for the same film  on  glass substrate. A similar effect was reported for a different system involving metallic nanostructured foam in a solution containing dye molecules.\cite{noginov-nanomaterials20} These observations point to an intimate interplay between FRET  and ET to the environment which results in a \textit{suppression} of concentration quenching for high molecule concentrations and, hence, in enhanced, rather than reduced, ensemble fluorescence.

In this paper, we present an analytical model, supported by numerical calculations, for FRET between the molecules in inhomogeneous and absorptive environment, e.g., near a metal interface, in the \textit{absence} of plasmonic enhancement effects. By extending the standard derivation of FRET rate to include an additional ET channel, we show that the effect of environment is to \textit{reduce} the F\"{o}rster radius by the factor $\sim (Q_{d}/q_{d})^{1/6}$, where $q_{d}$ is the donor's intrinsic quantum efficiency and  $Q_{d}$ is that in the presence of absorptive environment. We then demonstrate that, for a dye-doped film on top of a metal substrate,   the interplay between the increased donor's decay rate and reduced donor-acceptor FRET rate, both due to ET to the metal, leads to a highly non-trivial emission kinetics characterized by considerably slower, than for a grass substrate,  fluorescence decay rate that exhibits, for high dye concentrations, non-monotonic behavior with varying film thickness, consistent with the experiment.

\section{F\"{o}rster radius in inhomogeneous and absorptive environment}

We consider FRET from an excited donor to an acceptor  in  inhomogeneous and absorptive environment, e.g., near a metal-dielectric structure, which provides an additional ET channel competing with the donor-acceptor ET and radiation. We assume that all ET rates are much smaller than the optical frequency $\omega$ and adopt the standard approach to FRET but with slowly time-varying parameters. In the absorptive environment, the dipole moment of an excited donor decays with time as $\bm{p}_{d}(t)=\bm{p}_{d}e^{-\gamma t/2}$, where $\bm{p}_{d}=p_{d}\bm{n}_{d}$ is donor's initial dipole moment after the excitation ($\bm{n}_{d}$ is the dipole orientation) and $\gamma=\gamma_{d}+\gamma_{\rm et}$ is the decay rate. Here,  $\gamma_{\rm et}$ is ET rate from the donor to  absorptive environment and $\gamma_{d}= \gamma_{d}^{r}+\gamma_{d}^{nr}$ is the decay rate of an isolated donor comprised of radiative ($\gamma_{d}^{r}$) and nonratiative ($\gamma_{d}^{nr}$)  rates. The radiative decay rate has the standard form $\gamma_{d}^{r}=(4 \omega^{3}/3\hbar c^{3})|\bm{p}_{d}|^{2}$, while  $q_{d}=\gamma_{d}^{r}/\gamma_{d}$ is the  donor's intrinsic quantum efficiency. 

The F\"{o}rster ET represents  excitation of an acceptor  by the donor's  electric field $\bm{E}_{d}(\bm{r}_{a})$ at  the acceptor's position $\bm{r}_{a}$.  The  probability rate of such a process has the form \cite{ag-book}
\begin{equation}
W_{a}=\frac{2\pi}{\hbar}|\bm{p}_{a}\cdot \bm{E}_{d}(\bm{r}_{a})|^{2}f_{a}(\omega),
\end{equation}
where $\bm{p}_{a}=p_{a}\bm{n}_{a}$ is acceptor's dipole moment and $f_{a}(\omega)$ is its normalized spectral function that defines its absorption spectrum.  Using the relation between the spectral function and  the absorption cross section $\sigma_{a}(\omega)=(4\pi^{2}p_{a}^{2}\omega/3c)f_{a}(\omega)$, the FRET probability rate takes the form
\begin{equation}
\label{prob-rate}
W_{a}=\frac{3c}{2\pi\hbar\omega}\left |\bm{n}_{a}\cdot \bm{E}_{d}(\bm{r}_{a})\right |^{2}\sigma_{a}(\omega).
\end{equation}
Note that spectral linewidth of the absorption cross section which is determined by molecule's vibrational modes is typically much larger than its decay rates and, hence, the acceptor's absorption cross section is not significantly affected by the environment. The donor's electric field, which defines the transition matrix element, is related to its dipole moment as
\begin{equation}
\bm{E}_{d}(\bm{r}_{a})=\bm{D}(\omega,\bm{r}_{a},\bm{r}_{d})\bm{p}_{d}(t),
\end{equation}
where   $\bm{D}(\omega,\bm{r}_{a},\bm{r}_{d})$ is the electromagnetic (EM) Green function dyadic in the presence of environment. Accordingly, the probability rate (\ref{prob-rate}) takes the form,
\begin{equation}
\label{prob-rate-gen}
W_{a}(t)=\frac{3c}{2\pi\hbar\omega}\left |D_{ad}(\omega)\right |^{2}\sigma_{a}(\omega)|p_{d}(t)|^{2}.
\end{equation}
%
where $D_{ad}(\omega)=\bm{n}_{a} \bm{D}(\omega,\bm{r}_{a},\bm{r}_{d})\bm{n}_{d}$ is the donor-acceptor coupling. The latter can be presented as $D_{ad}=D_{ad}^{0}+D_{ad}^{\rm en}$, where $D_{ad}^{0}$ is direct dipole coupling and $D_{ad}^{\rm en}$ is determined by the environment.

In  the adiabatic approximation, the FRET probability rate (\ref{prob-rate-gen}) is time-dependent  as the donor, after initial excitation, is slowly, compared to the optical period, loosing its energy to the environment. To determine the F\"{o}rster radius in  inhomogeneous and absorptive environment, let us first recall  standard derivation of $R_{F}$ in the absence of it. In the latter case, $\bm{p}_{d}$ is time-independent and the donor-acceptor coupling contains only the direct dipole term $D_{ad}^{0}=\kappa_{ad}/ r_{ad}^{3}$,  where $r_{ad}$  is donor-acceptor  distance  and $\kappa_{ad}$  describes their dipoles' mutual orientation ($\kappa_{ad}^{2}=2/3$ for random orientations). In this case, the FRET probability rate (\ref{prob-rate-gen}) takes the  form 
\begin{equation}
\label{prob-rate-time}
W_{a}(r_{ad})=\frac{3c\kappa_{ad}^{2}}{2\pi\hbar\omega r_{ad}^{6}}\sigma_{a}(\omega)\left |p_{d}\right |^{2}.
\end{equation}
The F\"{o}rster radius $R_{F}$ is the characteristic distance defined by the condition \cite{ag-book} $\tau_{d}W_{a}(R_{F}) = 1$, where $\tau_{d}=\gamma_{d}^{-1}$ is isolated donor's fluorescence time, implying that, during time $\tau_{d}$,  same amount of energy is transferred from a donor to the acceptor situated at a distance $R_{F}$ as it is lost by an isolated donor to free-space fluorescence. For a donor with large spectral linewidth, this condition should be frequency-integrated with donor's normalized spectral function $f_{d}(\omega)$. Using the relation $\tau_{d}=q_{d}/\gamma_{d}^{r}=3q_{d}\hbar c^{3}/4p_{d}^{2}\omega^{3}$, we obtain the standard expression for the F\"{o}rster radius \cite{ag-book}
\begin{equation}
\label{rate-forster}
R_{F}^{6}= \frac{9c^{4}q_{d}\kappa_{ad}^{2}}{8\pi }\int\frac{d\omega}{\omega^{4}}f_{d}(\omega)\sigma_{a}(\omega).
\end{equation}
Accordingly, the normalized FRET rate $\gamma_{ad}$ takes the form 
\begin{equation}
\frac{\gamma_{ad}}{\gamma_{d}}=\left (\dfrac{R_{F}}{r_{ad}}\right )^{6},
\end{equation}
implying that,  for $r_{ad}<R_{F}$, FRET is the donor's dominant  energy loss channel.

In the presence of inhomogeneous and absorptive environment, the  F\"{o}rster radius should be modified since FRET is now accompanied by donor's energy loss to the environment. In the adiabatic approximation, the probability rate (\ref{prob-rate-gen})  evolves with  time following slow decay of the donor's dipole moment $ |p_{d}(t)|^{2}=|p_{d}|^{2}e^{-\gamma t}$. Note that the  donor-acceptor coupling $D_{ad}(\omega)$ is time-independent as it is determined by the EM Green function at the optical frequency $\omega$. Importantly, since $\gamma >\gamma_{d}$, the probability rate $W_{a}(t)$ can change substantially during time $\tau_{d}$. Therefore, when comparing to  $\gamma_{d}$, the time dependent probability rate $W_{a}(t)$ should be replaced with its \textit{time-averaged} value
\begin{equation}
\label{prob-rate-aver}
\bar{W_{a}}=\frac{1}{\tau_{d}}\int_{0}^{\infty} \! dt\,  W_{a}(t)=\frac{\gamma_{d}}{\gamma}\, \frac{3c}{2\pi\hbar\omega}\left |D_{ad}(\omega)\right |^{2}\sigma_{a}(\omega)|p_{d}|^{2},
\end{equation}
where the factor $\gamma_{d}/\gamma$ characterizes a reduction of average FRET probability rate due to a competing ET channel to the environment. Normalizing Eq.~(\ref{prob-rate-aver}) by donor's fluorescence rate $\gamma_{d}$ and integrating the result with normalized spectral function $f_{d}(\omega)$, we obtain  \textit{modified} FRET rate $\bar{\gamma}_{ad}$ as
\begin{equation}
\label{fret-rate-env}
\frac{\bar{\gamma}_{ad}}{\gamma_{d}}=\frac{9c^{4}Q_{d}}{8\pi }\int\frac{d\omega}{\omega^{4}}\left |D_{ad}(\omega)\right |^{2}f_{d}(\omega)\sigma_{a}(\omega),
\end{equation}
where $Q_{d}=q_{d}\gamma_{d}/\gamma=\gamma_{d}^{r}/\gamma$ is quantum efficiency of an isolated donor \textit{in}  absorptive environment. Here we assumed that the donor-acceptor spectral overlap is sufficiently narrow so that $Q_{d}$ does not significantly change within it.

The effect of environment on the FRET rate is twofold. First, the appearance of quantum efficiency $Q_{d}$ in place of the intrinsic quantum yield $q_{d}$ in Eq.~(\ref{fret-rate-env}) results in a \textit{reduction} of FRET rate at the same donor-acceptor distance $r_{ad}$. Second, the environment contribution to the donor-acceptor coupling $D_{ad}(\omega)=D_{ad}^{0}+D_{ad}^{\rm en}(\omega)$ can lead to its strong \textit{enhancement}, e.g., if the donor-acceptor  pair is  situated near a plasmonic system, in which case the overall effect is determined by the system geometry.\cite{pustovit-prb11} In the rest of the paper, we consider systems with \textit{no} plasmonic enhancement in the relevant frequency range and sufficiently close donor-acceptor distances. In this case, as we show in the Appendix, we can disregard the environment-induced coupling $D_{ad}^{\rm en}$ and include the direct coupling $D_{ad}^{0}$  only in our further analysis.

Keeping the direct coupling $D_{ad}^{0}=\kappa_{ad}/ R_{ad}^{3}$ in Eq.~(\ref{fret-rate-env}) and comparing  to Eq.~(\ref{rate-forster}), we obtain the modified FRET rate as 

\begin{equation}
\label{rate-forster-metal}
\frac{\bar{\gamma}_{ad}}{\gamma_{d}}=\left (\dfrac{\bar{R}_{F}}{R_{ad}}\right )^{6},
\end{equation}
where $\bar{R}_{F}$ is  \textit{modified} F\"{o}rster radius defined as
\begin{equation}
\label{forster-ratio}
\bar{R}_{F}^{6}=\frac{Q_{d}}{q_{d}}R_{F}^{6} =\frac{\gamma_{d}}{\gamma}R_{F}^{6}.
\end{equation}
The reduction of F\"{o}rster radius implies that, in the presence of absorptive environment, a donor should be situated closer to an acceptor in order to transfer to it, during time $\tau_{d}$,  the same  amount of energy that is lost to free-space fluorescence.  

\section{FRET and emission kinetics for large ensembles of molecules in inhomogeneous and absorptive environment}

We now turn to fluorescence from large ensembles of molecules in absorptive environment. We assume that the Stokes shift is small enough to provide sufficient overlap between the molecules' absorption and emission bands. In this case, fluorescence by an excited molecule (donor) is accompanied by FRET to nearby molecules (acceptors) which, for high molecule concentrations, leads to concentration quenching of the ensemble fluorescence. On the other hand, fluorescence of a molecule placed in  absorptive environment, e.g., near a metal interface, is quenched due to ET to the environment. In this case, the emission kinetics is described by fast exponential decay of the fluorescence intensity $I(t)=I_{0}e^{-\gamma t}$, where $I_{0}$ is the initial intensity and $\gamma=\gamma_{d}+\gamma_{\rm et}$ is individual molecule's decay rate.\cite{silbey-review} However, for dye-doped polymer films deposited on top of metal substrate, a substantially \textit{slower} emission kinetics was observed for \textit{high molecule concentrations} as compared to same system on glass substrate.\cite{noginov-josa21} Furthermore, non-monotonic behavior with varying film thickness was observed for the effective fluorescence decay rate, in sharp contrast to nearly constant rate for the same system on glass substrate, implying an intimate interplay between the metal quenching and concentration quenching which we  address below.

To illustrate the underlying mechanism of concentration quenching, consider light emission by an excited donor in the presence of acceptors with high concentration $n_{a}$. In non-absorptive environment, the emission kinetics is described by  exponential decay $I(t)=I_{0}e^{-(\gamma_{d}+\gamma_{F})t}$, where
\begin{equation}
\gamma_{F}= \sum_{a}  \gamma_{ad}  =\gamma_{d}\sum_{a}\left (\frac{ R_{F}}{r_{ad}}\right )^{6}
\end{equation}
is  decay rate  associated with FRET from a donor to surrounding acceptors. If acceptors are distributed uniformly within some confined region,  then a standard averaging procedure over their positions yields \cite{huber-prb85,parkinson-prb07,krenn-nl08}   
\begin{align}
\label{kinetics}
\langle e^{-\gamma_{F}t}\rangle
&=
\langle\prod_{a}e^{-\gamma_{d}t( R_{F}/r_{ad})^{6}}\rangle 
\\
&= \exp\left [-n_{a}\int d\bm{r}\left [ 1-e^{-\gamma_{d}t( R_{F}/r)^{6}}\right ]\right ],
\nonumber
\end{align}
where integration takes place over donor-acceptor radius-vectors $\bm{r}$. In general, for molecules distributed in a confined region, the kinetics depends on system geometry. However,  if the region size is sufficiently large, the fraction of donors near its boundary is relatively small and the integration can be extended over the entire volume. In this case, one obtains an analytical expression for the emission kinetics \cite{huber-prb85,parkinson-prb07,krenn-nl08}
\begin{equation}
\label{kinetics-forster}
I(t)=I_{0}e^{-\gamma_{d}t-N_{a}\sqrt{\pi\gamma_{d}t}},
\end{equation}
where  $N_{a}= n_{a}V_{F}$  is number of acceptors within F\"{o}rster's volume $V_{F}=4\pi R_{F}^{3}/3$. With increasing acceptor concentration, as $N_{a}$ exceeds several molecules in Forster's volume,  the second term in the exponent becomes dominant and causes strong reduction of fluorescence intensity (concentration quenching). Note that if the acceptors are distributed within a plane,  the $t^{1/2}$ behavior in the exponent  is replaced with $t^{1/3}$ one.\cite{parkinson-prb07,krenn-nl08}

In the presence of absorptive environment, the emission kinetics is described by modified fluorescence intensity  $I(t)=I_{0}e^{-(\gamma+\bar{\gamma}_{F})t}$, where the decay rate $\gamma=\gamma_{d}+\gamma_{\rm et}$ now includes ET to the environment while the FRET rate $\bar{\gamma}_{F}=\sum_{a}  \bar{\gamma}_{ad}$ is  related to the modified F\"{o}rster's radius $\bar{R}_{F}$ as
\begin{equation}
\bar{\gamma}_{F}
 =\gamma_{d}\sum_{a}\left (\frac{ \bar{R}_{F}}{r_{ad}}\right )^{6},
\end{equation}
where $\bar{R}_{F}$ is defined by Eq.~(\ref{forster-ratio}). The averaging over the acceptors' positions can also be performed using Eq.~(\ref{kinetics}). Note that, since the modified F\"{o}rster's radius is reduced near the metal interface, the boundary effects are unimportant and, in a good approximation, the integration in Eq.~(\ref{forster-ratio}) can be extended to the entire volume (see Appendix for details). Finally, the donnor's emission kinetics is described by a similar expression but with modified parameters
\begin{equation}
\label{kinetics-forster-mod}
I(t)=I_{0}e^{-\gamma t-\bar{N}_{a}\sqrt{\pi\gamma_{d}t}}.
\end{equation}
where $\bar{N}_{a}= n_{a}\bar{V}_{F}$ is  the number of acceptors within \textit{modified}  F\"{o}rster's volume $\bar{V}_{F}=4\pi \bar{R}_{F}^{3}/3$. Importantly, the  absorptive  environment exerts \textit{opposite} effects on the two terms in the exponent of  Eq.~(\ref{kinetics-forster-mod}). While the first term, describing donor's energy loss to the environment and radiation, is \textit{enhanced} by the large factor $\gamma/\gamma_{d}$ as compared to that in Eq.~(\ref{kinetics-forster}), the second term, describing the average FRET to the acceptors, is \textit{reduced} by the factor
\begin{equation}
\label{forster-volume-mod}
\frac{\bar{N}_{a}}{N_{a}}=\frac{\bar{V}_{F}}{V_{F}}=\sqrt{\frac{\gamma_{d}}{\gamma}},
 \end{equation} 
due to decrease of the F\"{o}rster radius [see Eq.~(\ref{forster-ratio})].  In fact, as we demonstrate in the numerical calculations below, these two quenching mechanisms compete against each other: an increase in environment-induced non-radiative decay rate $\gamma_{\rm et}$ results in a suppression of concentration quenching.

\section{Discussion and numerical results}

%
\begin{figure}[b]
\vspace{2mm}
\centering
\includegraphics[width=1.\columnwidth]{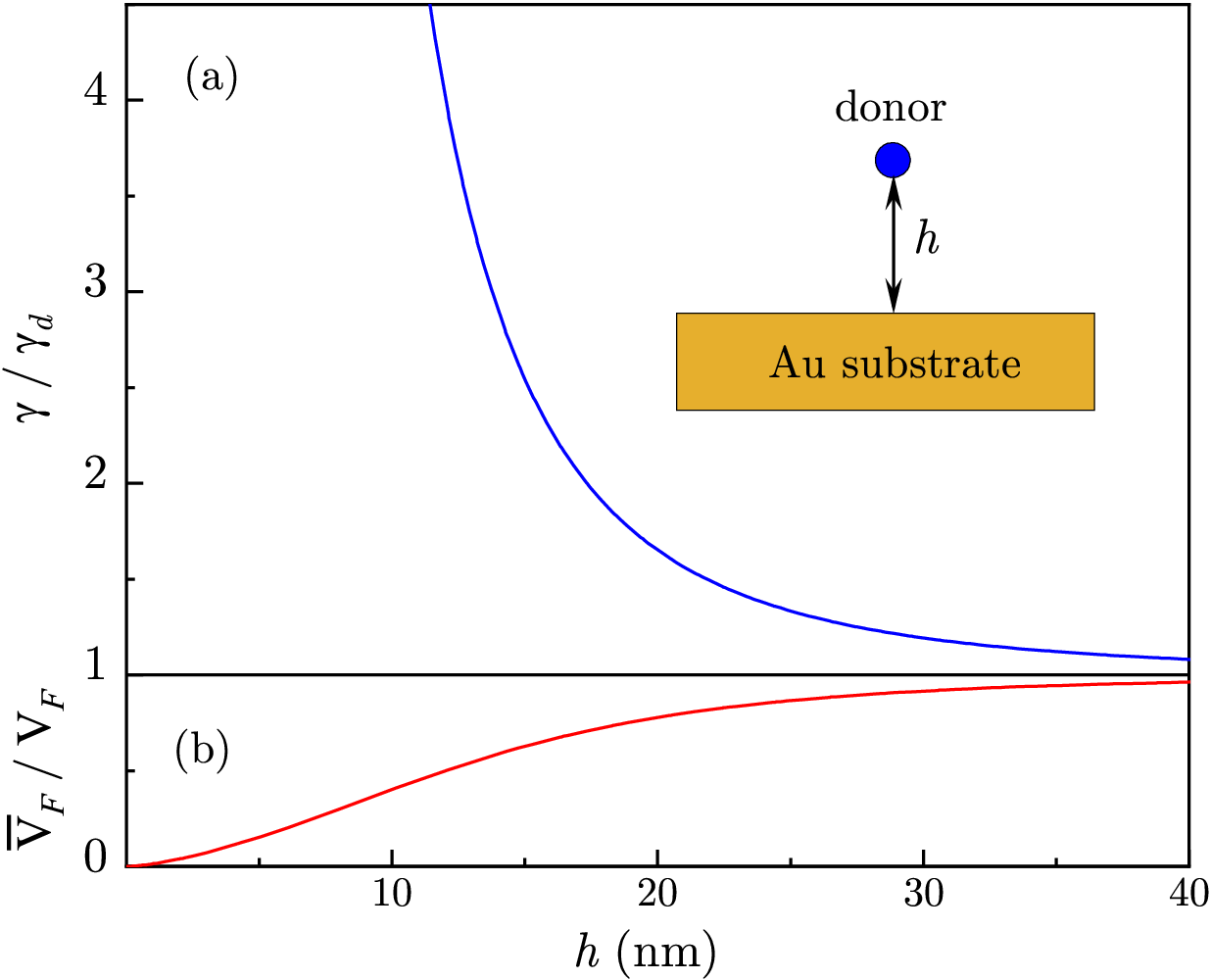}

%
\caption{\label{fig1} Normalized donor decay rate (a) and modified  F\"{o}rster volume (b) are plotted against donor's distance to a flat Au surface.
 }
\vspace{-4mm}
\end{figure}
%

Here we present the results of numerical calculations for dye-doped polymeric films (HITC:PMMA) of thickness $L$ deposited on gold or glass substrates, which is similar to setup used in recent experiments.\cite{noginov-josa21} The HITC dyes absorption and emission spectra are centered at 762 nm and 772 nm, respectively, with intrinsic quantum yield $q_{d}=0.3$, providing efficient FRET at the F\"{o}rster radius  $R_{F}\approx 5$ nm, so the value $N_{a}=1$ corresponds to concentration $n_{a}\approx 1.7$ g/l. We used experimental gold dielectric function in all calculations. The ET rate $\gamma_{\rm et}$ to the metal interface  and the coupling $D_{ad}(\omega)$ were computed using EM Green function in the presence of planar metal surface,\cite{silbey-review} while the averaging over acceptors' positions in Eq.~(\ref{kinetics}) was performed within the film volume. The fluorescence intensity for the film $I(L,t)$ was calculated by averaging out the intensity $I(h,t)$ for an individual molecule, situated  at a distance $h$ above the interface, over the film thickness $L$ as $I(L,t)=L^{-1}\int_{0}^{L\!}dh\, I(h,t)$. The effective fluorescence decay rate $\Gamma$ was approximated as $\Gamma^{-1}=\int_{0}^{\infty}dt\, I(L,t)/ I(L,0)$, which is approximately equivalent to  fitting the normalized intensity by a simple exponential decay $e^{-\Gamma t}$ in the experiment.\cite{noginov-nanophot21,noginov-josa19,noginov-josa21,noginov-nanomaterials20} The full details of calculations are presented in the Appendix.

In Fig.~\ref{fig1}, we plot the the individual donor's decay rate $\gamma=\gamma_{d}+\gamma_{\rm et}$ and modified F\"{o}rster volume $\bar{V}_{F}$ for a donor situated at a distance $h$ above the metal interface. With decreasing $h$, the decay rate $\gamma$ sharply increases as the ET rate  behaves as $\gamma_{\rm et} \propto h^{-3}$ close to the interface.\cite{silbey-review} In contrast, the modified F\"{o}rster volume $\bar{V}_{F}$ decreases according to Eq.~(\ref{forster-volume-mod}) and, correspondingly,  behaves as $\bar{V}_{F}\propto h^{3/2}$ close to the metal interface. 

%
\begin{figure}[bt]
\vspace{2mm}
\centering
\includegraphics[width=1.\columnwidth]{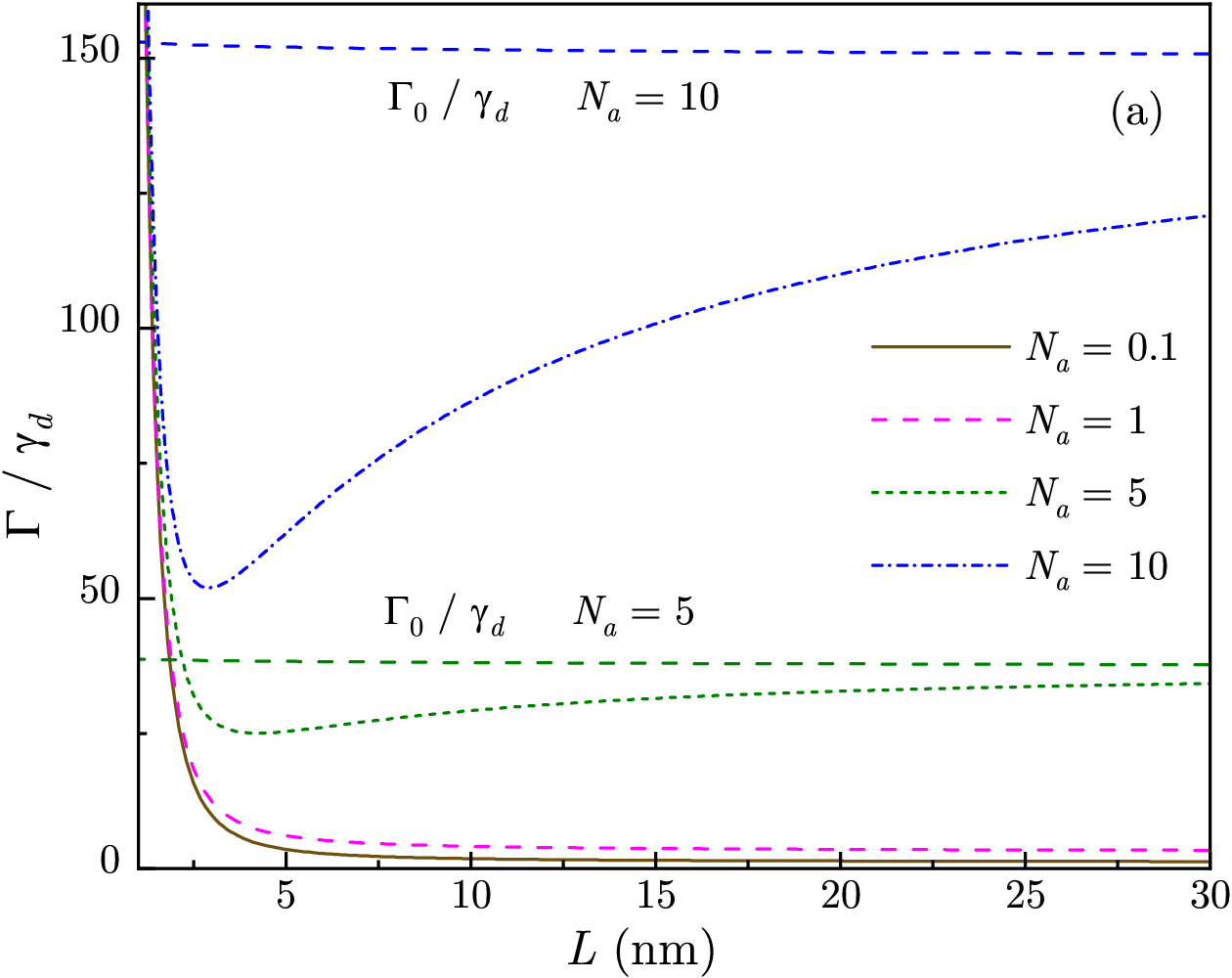}

\vspace{3mm}

\includegraphics[width=1.\columnwidth]{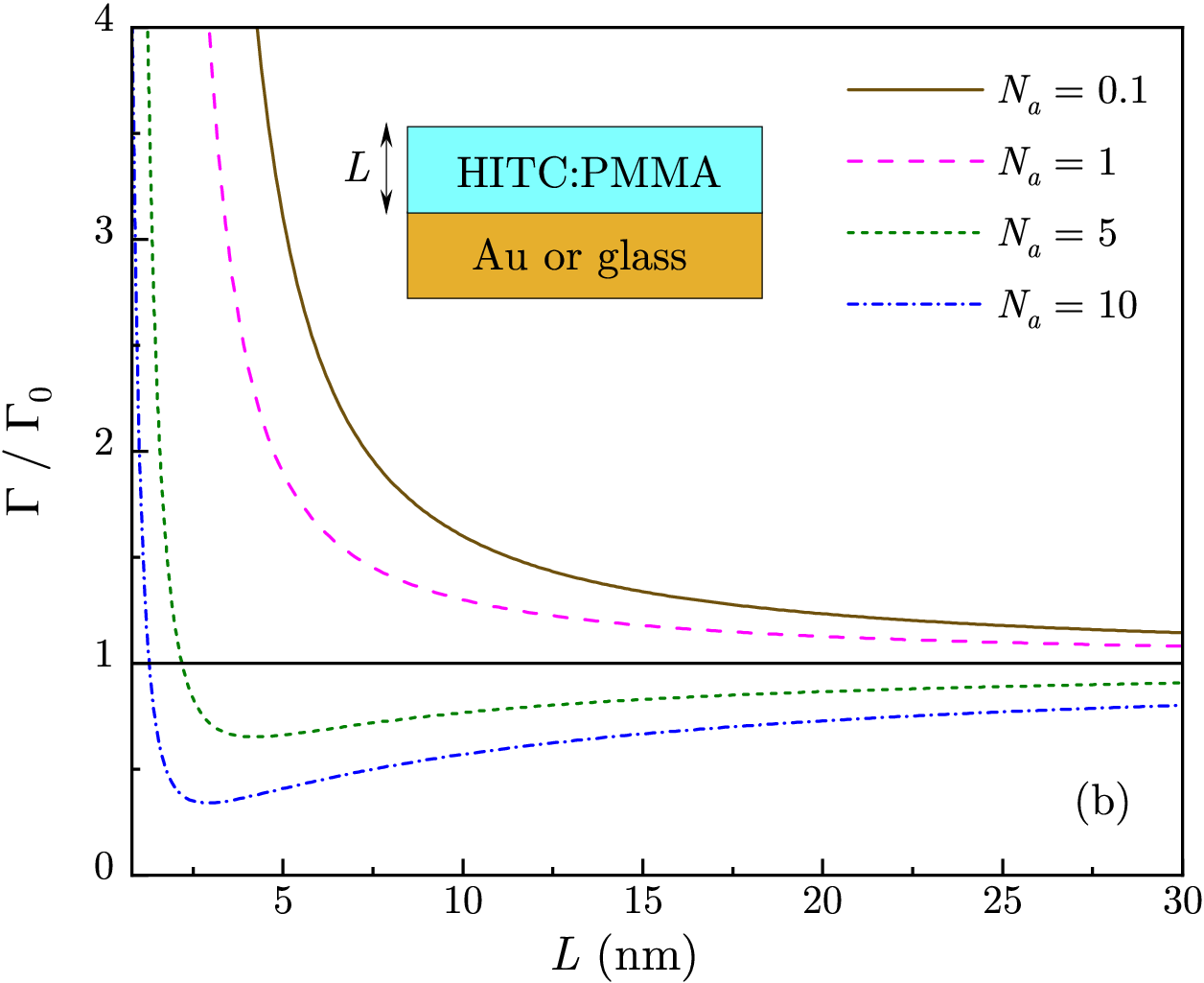}
\caption{\label{fig2} (a) Effective fluorescence decay rate is shown versus thickness $L$ of dye-doped polymer film on top of Au substrate ($\Gamma$) and grass substrate ($\Gamma_{0}$)  for several values of dye concentration. (b) The ratio $\Gamma/\Gamma_{0}$ is shown for the same parameter values. Inset: Schematics of dye-doped polymer film on top of Au or glass substrate.
 }
\vspace{-4mm}
\end{figure}
%

In Fig.~\ref{fig2}, we demonstrate the interplay between the metal quenching and concentration quenching.  Fig.~\ref{fig2}(a) shows the effective fluorescence decay rate $\Gamma$ with varying film thickness $L$ for various molecule concentrations. For low concentrations $N_{a}\leq 1$, i.e., less than one molecule within the  F\"{o}rster volume $V_{F}$, the dominant decay mechanism is ET to the metal, which leads to fluorescence quenching for small $L$, while for larger $L$, the decay rate is close to $\gamma_{d}$ as the molecules are situated, on average, further away from the metal. For higher concentrations, $\Gamma$ increases substantially for thicker films signaling the dominant role of concentration quenching away from the metal. Remarkably, the dependence of effective decay rate $\Gamma$ on $L$ is \textit{non-monotonic}, in contrast to rate $\Gamma_{0}$ for the same system on glass, shown for comparison in Fig.~\ref{fig2}(a), which stays nearly constant with varying $L$ indicating that it is not very sensitive to the boundary effects.  In Fig.~\ref{fig2}(b), we plot the effective decay rate for the metal substrate $\Gamma$ normalized by that for the glass substrate $\Gamma_{0}$ with varying $L$ for the same dye concentrations. For high concentrations, at which several dyes are confined within the F\"{o}rster volume $V_{F}$, the effective decay rate $\Gamma$ is substantially reduced as compared to $\Gamma_{0}$, except for small film thicknesses $L$ at which fluorescence is quenched by the metal. Such an interplay between the metal quenching and concentration quenching leads, for high concentrations, to a \textit{slower} fluorescence decay on the metal substrate as compared to glass, consistent with the experiment.\cite{noginov-josa21,noginov-nanomaterials20}

%
\begin{figure}[tb]
\vspace{2mm}
\centering
\includegraphics[width=1.\columnwidth]{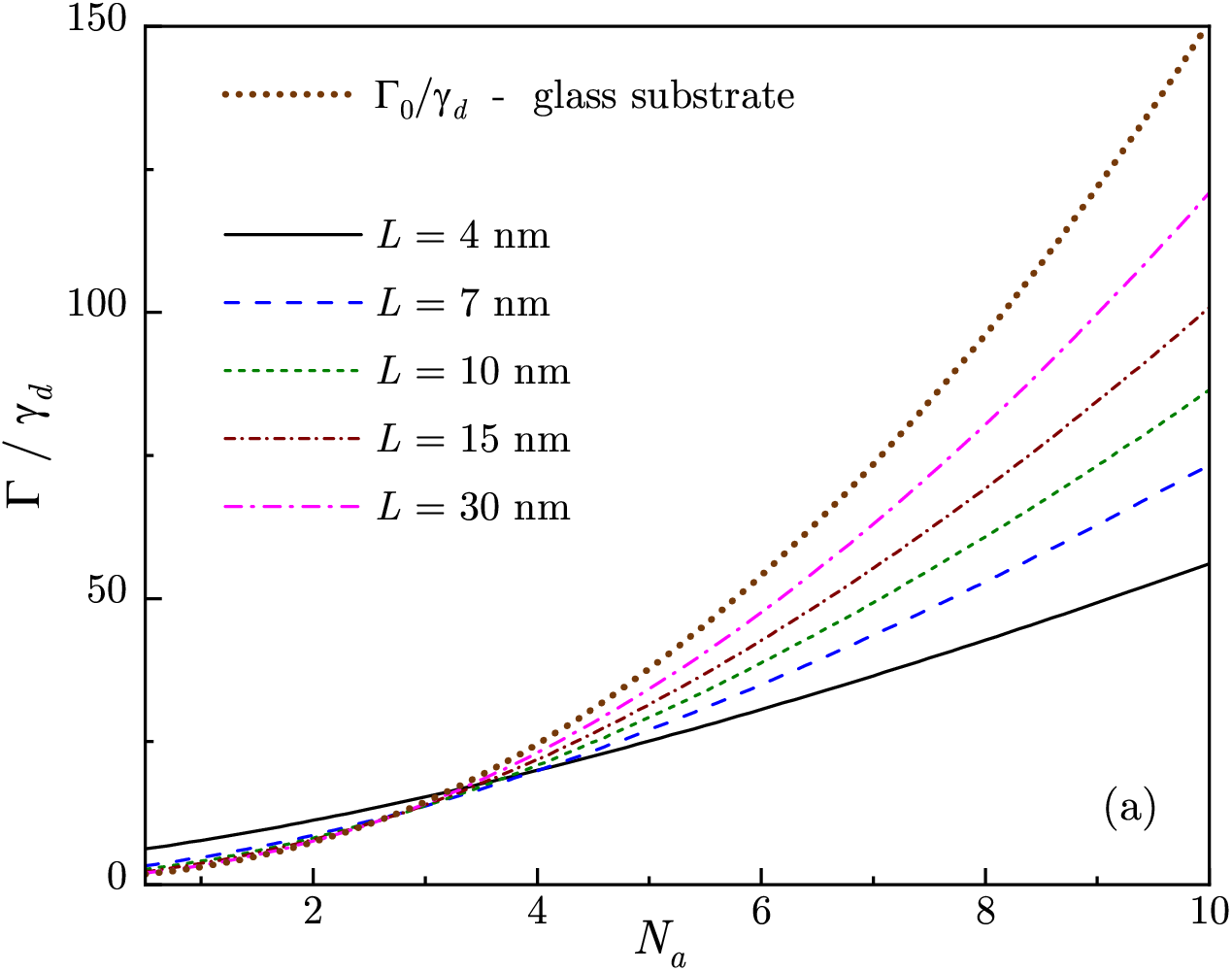}

\vspace{3mm}

\includegraphics[width=1.\columnwidth]{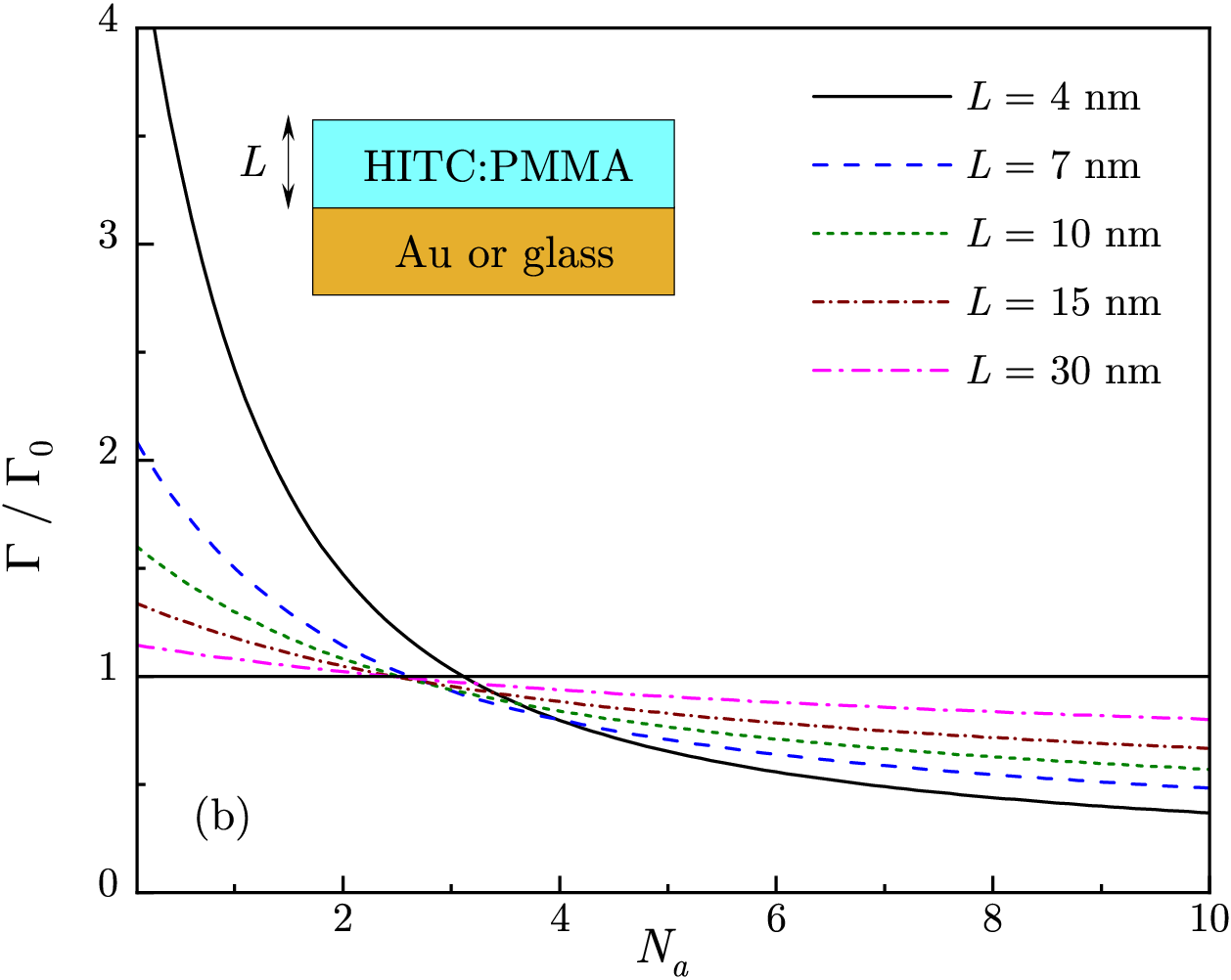}
\caption{\label{fig3} (a) Effective fluorescence decay rate is shown versus dye concentration for several values of thickness $L$ of dye-doped polymer film on top of Au substrate ($\Gamma$) and grass substrate ($\Gamma_{0}$). (b) The ratio $\Gamma/\Gamma_{0}$ is shown for the same parameter values. Inset: Schematics of dye-doped polymer film on top of Au or glass substrate.
 }
\vspace{-4mm}
\end{figure}
%

In Fig.~\ref{fig3} we illustrate the suppression of concentration quenching by plotting $\Gamma$ and $\Gamma_{0}$ with increasing molecule concentration for different film thicknesses $L$. For each film thickness $L$, the rate $\Gamma$ is slower than $\Gamma_{0}$ for concentrations exceeding several molecules within the F\"{o}rster volume [see Fig.~\ref{fig3}(a)]. The effect of metal substrate is stronger in thinner films both for low and high concentrations, as shown in Fig.~\ref{fig3}(b). For low concentrations, as expected, the ratio $\Gamma/\Gamma_{0}$ is larger for thinner  films, where the molecules are closer,  on average, to the metal. On the other hand, for high concentrations, $\Gamma/\Gamma_{0}$ is now smaller in thinner films due to a stronger suppression of concentration quenching. This unexpected behavior reflects the intimate interplay between FRET and ET to the absorptive environment.

\section{conclusions}

In summary, we have developed an analytical model, supported by numerical calculations, for  F\"{o}rster resonance energy transfer (FRET) between a donor and an acceptor placed in inhomogeneous and absorptive environment, e.g., near a metal-dielectric structure. By extending the standard approach to FRET to include the additional energy transfer (ET) channel, we found that, in the absence of plasmon resonances, the F\"{o}rster radius can be significantly reduced due to a competing ET process from a donor to the environment. We studied the implications of F\"{o}rster radius reduction for fluorescence from large ensemble of molecules, whose emission kinetics is dominated by  concentration quenching. Specifically, we performed numerical calculations for dye-doped polymer films of various thicknesses deposited on top of  metallic and glass substrates. We found that, for high molecule concentrations, the appearance of competing ET channel to the metal results in slowing down of the emission kinetics, in sharp contrast to single-molecule fluorescence kinetics. Furtheremore, for high molecule concentrations, the effective fluorescence decay rate exhibits non-monotonic behavior with varying film thickness, consistent with the experiment. Unexpectedly, such effects are stronger in thinner films with smaller average molecule distance the metal interface, indicating a non-trivial interplay between the metal quenching and concentration quenching mechanisms. 

\acknowledgements
This work was supported in part by the National Science Foundation grants DMR-2000170, DMR-2301350, and NSF-PREM-2423854.


\appendix*

\section{}

Here we describe details of calculations of the decay rate $\gamma$ and modified F\"{o}rster radius $\bar{R}_{F}$ for a donor-acceptor pair situated near planar metallic interface. In the system we consider the molecules' distances to the interface are much smaller that the radiation wavelength and so we can use the near-field limit of the EM Green function.\cite{silbey-review} For a donor situated at a distance $h$ from the interface, the decay rate $\gamma=\gamma_{d}+\gamma_{\rm et}$ normalized by the free-space  fluorescence decay rate  $\gamma_{d}$ has the form
\begin{equation}
\label{donor-gamma}
\frac{\gamma}{\gamma_{d}}=1+\frac{3\kappa q_{d}}{8(kh)^{3}}\, \text{Im}\frac{\varepsilon(\omega_{d})-1}{\varepsilon(\omega_{d})+1},
\end{equation}
where $k$ is the wave vector,  $\kappa$ is the dipole orientation factor ($\kappa_{\perp}=1$ and $\kappa_{\parallel}=1/2$), and $\varepsilon(\omega_{d})$ is the metal dielectric function at the donor's emission frequency $\omega_{d}$. The medium dielectric constant $\varepsilon_{\rm m}$ is included in the numerical calculations by replacement $\varepsilon(\omega_{d})\rightarrow \varepsilon(\omega_{d})/\varepsilon_{\rm m}$.

\begin{figure}[b]
\vspace{2mm}
\centering
\includegraphics[width=0.65\columnwidth]{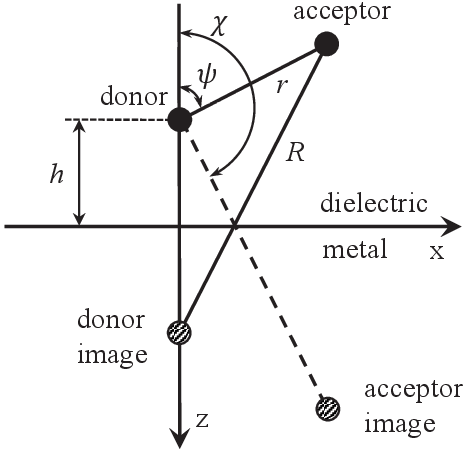}

%
\caption{\label{fig-app1} Schematics of interacting donor and acceptor placed above planar metal interface.
 }
\end{figure}

Near the metal interface, the donor-acceptor coupling $D_{ad}$ includes  direct dipole interaction and interactions with image dipoles (see Fig.~\ref{fig-app1}). In the near-field limit, the coupling has the form $D_{ad}=\bm{\mu}_{d}\bm{V}\bm{\mu}_{a}$, where $\bm{\mu}_{d}$ and $\bm{\mu}_{a}$ are donor's and acceptor's dipole moment vectors, respectively, and the dyadic $\bm{V}$ has the matrix form \cite{silbey-cpl95}
\begin{align}
\label{v-matrix}
\bm{V} =
&\dfrac{1}{2r^3}
\begin{pmatrix}
3\cos 2 \psi -1 & 0 & 3 \sin 2 \psi\\
0 & 2 & 0\\
3 \sin 2 \psi & 0 & -3\cos 2 \psi -1
\end{pmatrix}
\\
&-\dfrac{1}{2R^3} \dfrac{\epsilon (\omega) - 1}{\epsilon (\omega) + 1}
\begin{pmatrix}
3\cos 2 \chi -1 & 0 & 3 \sin 2 \chi \\
0 & 2 & 0\\
-3 \sin 2 \chi & 0 & 3\cos 2 \chi +1
\end{pmatrix},
\nonumber
 \end{align}
where the matrix parameters  are given by (see Fig.~\ref{fig-app1})
\begin{align}
&R = \sqrt{r^{2} + 4 h^{2} + 4h r\cos \psi},
\nonumber\\
&\chi = \pi - \tan ^{-1} \left( \frac{r\sin \psi}{2h+r\cos \psi}\right).
\end{align}
The first and second terms in Eq.~(\ref{v-matrix}) describe, respectively, the direct dipole coupling and coupling via the image dipoles. The modified FRET rate is given by 
\begin{equation}
\label{fret-rate-plane}
\frac{\bar{\gamma}_{ad}}{\gamma_{d}}=\frac{9c^{4}Q_{d}}{8\pi }\left |D_{ad}(\omega_{o})\right |^{2}\int\frac{d\omega}{\omega^{4}}f_{d}(\omega)\sigma_{a}(\omega),
\end{equation}
where we $\left |D_{ad}(\omega_{o})\right |^{2}$ is taken at the central frequency $\omega_{o}$ of the donor and acceptor spectral overlap. The modified F\"{o}rster radius $r=\bar{R}_{F}$ is determined from the condition $\bar{\gamma}_{ad}/\gamma_{d}=1$. Averaging over the molecules' dipole orientations and using  definition of the F\"{o}rster radius $R_{F}$ in the absence of metal, 
\begin{equation}
\label{rate-forster-av}
R_{F}^{6}= \frac{3c^{4}q_{d}}{4\pi }\int\frac{d\omega}{\omega^{4}}f_{d}(\omega)\sigma_{a}(\omega),
\end{equation}
the  F\"{o}rster radius condition takes the form
\begin{align}
\label{forster-angle}
\frac{\gamma}{\gamma_{d}}  
&=\frac{R_{F}^{6}}{\bar{R}_{F}^{6}}
\Biggl [1 + \dfrac{\bar{R}_{F}^{6}}{4\bar{R}^6} \dfrac{|\epsilon(\omega_{o}) - 1|^{2}}{|\epsilon(\omega_{o}) - 1|^{2}} \left[ 1 + 3\cos4\bar{\chi}\right] 
\nonumber\\
&+\frac{3\bar{R}_{F}^{3}}{2\bar{R}^3} \dfrac{|\epsilon(\omega_{o})| ^ 2 - 1} {|\epsilon (\omega_{o}) + 1|^{2}}\left[ 3 (\cos 2 \psi +\cos 2\bar{\chi}) -2\right] \Biggr ],
 \end{align}
where we introduced  $\bar{R} = \sqrt{\bar{R}_{F}^{2} + 4 h^{2} + 4h \bar{R}_{F}\cos \psi}$ and $\bar{\chi} = \pi - \tan ^{-1} \left[ \bar{R}_{F}\sin \psi/(2h+\bar{R}_{F}\cos \psi)\right]$. The first term in the right hand side is the direct dipole coupling while the second and third terms describe the effect of image dipoles. Solutions of Eq.~(\ref{forster-angle}) define the modified F\"{o}rster radius $\bar{R}_{F}(h,\psi)$, which depends on the donor's distance  $h$ to the interface and donor-acceptor mutual orientation angle $\psi$ (see Fig.~\ref{fig-app1}). Note that  orientation dependence of the modified F\"{o}rster radius comes solely from the image dipoles.

%
%

%
\begin{figure}[tb]
\vspace{2mm}
\centering
\includegraphics[width=1.\columnwidth]{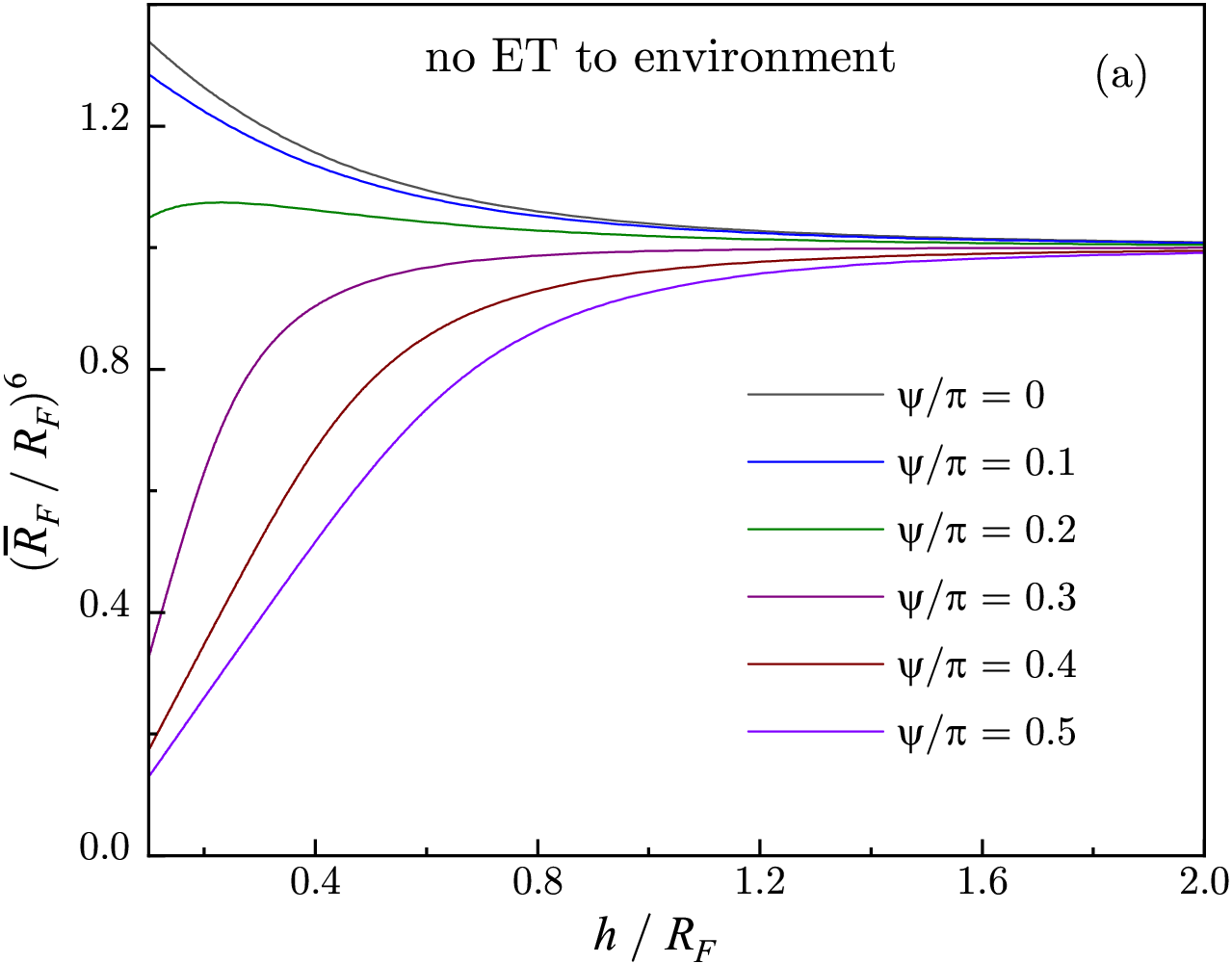}

\vspace{3mm}

\includegraphics[width=1.\columnwidth]{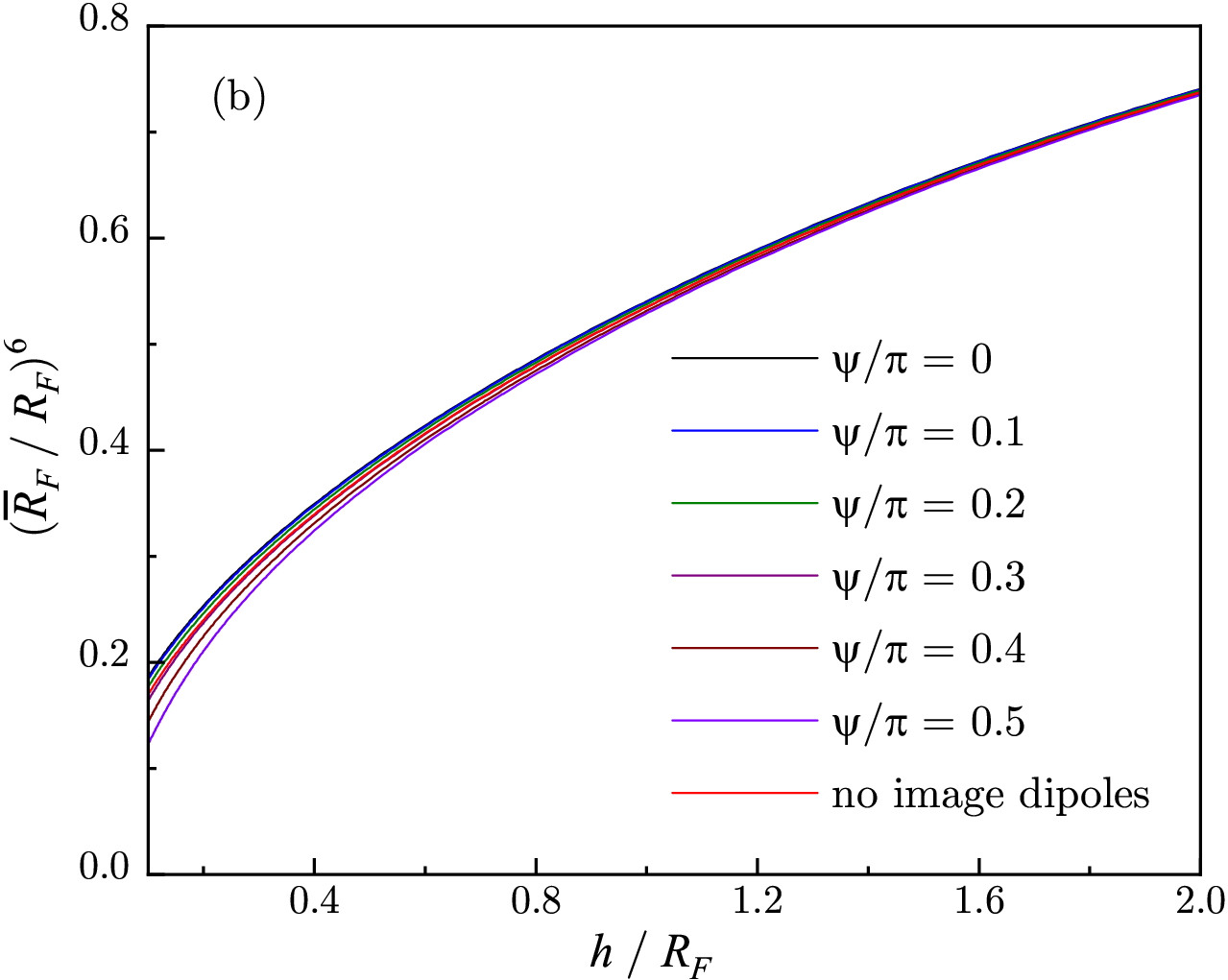}
\caption{\label{fig-app2} Modified F\"{o}rster radius $\bar{R}_{F}$ is shown versus donor's distance $h$ to the interface without (a) and with (b) ET to the environment for various donor-acceptor orientation angles $\psi$.
 }
\vspace{-4mm}
\end{figure}
%
%

In Fig.~\ref{fig-app2} we plot the modified F\"{o}rster radius $\bar{R}_{F}$ versus donor's distance $h$ to the interface for various values of angle $\psi$. In the absence of ET to the environment, i.e., $\gamma_{\rm et}=0$, the modified F\"{o}rster radius exhibits strong dependence on $\psi$, as illustrated in Fig.~\ref{fig-app2}(a), pointing at significant role of the image dipoles for close donor distances to the interface. However, the full solution of Eq.~(\ref{forster-angle}) incorporating ET to the metal shows that   $\bar{R}_{F}(h,\psi)$ is virtually independent of  $\psi$, as illustrated in Fig.~\ref{fig-app2}(b). Furthermore, the modified  F\"{o}rster radius calculated using Eq.~(\ref{forster-angle}) is very close to the dependence  $\bar{R}_{F}^{6} =(\gamma_{d}/\gamma)R_{F}^{6}$, i.e., \textit{without} the image dipoles effect [see Eq.~(\ref{forster-ratio})]. The reason for  weak orientation dependence of $\bar{R}_{F}$ in the presence of ET to the environment is a reduction of the ratio $\bar{R}_{F}/h$ and, hence, of the ratio $\bar{R}_{F}/\bar{R}$ in Eq.~(\ref{forster-angle}), which renders the second and third terms to be small as compared to the first direct dipole coupling term. Since the  image dipoles play no significant role for larger distances $h>R_{F}$ as well, we conclude that the simple analytical model  for modified FRET radius (\ref{forster-ratio}) and for  emission kinetics (\ref{kinetics-forster-mod}) in the presence of inhomogeneous absorptive environment is valid in a wide range of system parameters.



\begin{thebibliography}{99}

\bibitem{forster-ap48} T. F\"{o}rster, Ann. Phys. \textbf{437}, 55 (1948).
\bibitem{dexter-jcp53}D. L. Dexter, J. Chem. Phys. \textbf{21}, 836 (1953).

\bibitem{andrews-book} \textit{Resonance Energy Transfer}, edited by D. L. Andrews and A. A. Demidov (Wiley, New York, 1999).

\bibitem{ag-book} V. M. Agranovich and M. D. Galanin, \textit{Electronic Excitation Energy Transfer in Condensed 
Matter} (North‐Holland, Amsterdam, 1983).

\bibitem{lakowicz-book}J. R. Lakowicz, \textit{Principles of Fluorescence Spectroscopy} (Springer, New York, 2006).

%
%



\bibitem{deniz-pnas00} A. A. Deniz, T. A. Laurence, G. S. Beligere, M. Dahan,
A. B. Martin, D. S. Chemla, P. E. Dawson, P. G. Schultz,
and S. Weiss, Proc. Natl. Acad. Sci. USA \textbf{97}, 5179 (2000).

\bibitem{lipman-science03}
E. A. Lipman, B. Schuler, O. Bakajin, and W. A. Eaton,
Science \textbf{301}, 1233 (2003).

\bibitem{selvin-naturesb00}
P. R. Selvin, Nature Structural Biology \textbf{7}, 730 (2000).

\bibitem{sekar-jcb03} R. B. Sekar and A. Periasamy, J. Cell Biol. 160, 629 (2003).

\bibitem{gonzalez-bj95}
J. Gonzalez and R. Tsien, Biophys. J. 69, 1272 (1995).

\bibitem{medintz-naturemat03} I.L. Medintz, A.R. Clapp, H. Mattoussi, E.R. Goldman, and J.M. Mauro,
Nat. Mater. \textbf{2}, 630 (2003). 


\bibitem{andrews-lp11}
D. L. Andrews, C. Curutchet, and G. D. Scholes, Laser
Photonics Rev. 5, 114 (2011).



\bibitem{hopmeier-prl99} M. Hopmeier, W. Guss, M. Deussen, E. O. G\"{o}bel, R. F. Mahrt, Phys. Rev. Lett. \textbf{82}, 4118 (1999).

\bibitem{andrew-science00} P. Andrew and W. L. Barnes, Science \textbf{290}, 785 (2000). 

\bibitem{finlayson-cpl01} C. E. Finlayson, D. S. Ginger, N. C. Greenham, Chem. Phys. Lett. 338, 83-87 (2001).



\bibitem{leitner-cpl88} A. Leitner and H. Reinisch, Chem. Phys. Lett. \textbf{146}, 320-324 (1988). 

\bibitem{lakowicz-jf03} J. R. Lakowicz, J.  Ku\`{s}ba,  Y. Shen,  J.  Malicka,  S.  D’Auria,  Z. Gryczynski, and I.  Gryczynski, J. Fluoresc. \textbf{13}, 69 (2003).

\bibitem{andrew-science04} P. Andrew and W. L. Barnes,  Science \textbf{306}, 1002 (2004). 

\bibitem{lakowicz-jpcc07-1} J. Zhang, Y. Fu, and J. R. Lakowicz, J. Phys. Chem. C \textbf{111}, 50 (2007).

\bibitem{lakowicz-jpcc07-2} J. Zhang, Y. Fu, M. H. Chowdhury, and J. R. Lakowicz, J. Phys. Chem. C \textbf{111}, 11784 (2007).

\bibitem{krenn-nl08} F. Reil, U. Hohenester, J. R. Krenn, and A. Leitner, Nano Lett. 8, 4128 (2008).

\bibitem{rogach-apl08} V. K. Komarala, A. L. Bradley, Y. P. Rakovich, S. J. Byrne, Y. K. Gunko, and A. L. Rogach, Appl. Phys. Lett. \textbf{93}, 123102 (2008).

\bibitem{bodreau-nl09} M. Lessard-Viger, M. Rioux, L. Rainville, and D. Boudreau, 
Nano Lett. \textbf{9}, 3066 (2009).

\bibitem{yang-oe09} K. Y. Yang, K. C. Choi, and C. W. Ahn, Opt. Express \textbf{17}, 11495 (2009).

\bibitem{an-oe10}K. H. An, M. Shtein, and K. P. Pipe, Opt. Express \textbf{18}, 4041 (2010).

\bibitem{lunz-nl11} M. Lunz, V. A. Gerard, Y. K. Gun’ko, V. Lesnyak, N. Gaponik, A. S. Susha, A. L. Rogach , and A. L. Bradley,
Nano Lett. \textbf{11}, 3341 (2011).

\bibitem{zhao-jpcc12} L. Zhao, T. Ming, L. Shao, H. Chen, and J. Wang, 
J. Phys. Chem. C \textbf{116}, 8287 (2012).

\bibitem{west-jpcc12} R. G. West and S. M. Sadeghi, J. Phys. Chem. C \textbf{116}, 20496 (2012).

\bibitem{lunz-jpcc12} M. Lunz, X. Zhang, V. A. Gerard, Y. K. Gunko, V. Lesnyak, N. Gaponik, A. S. Susha, A. L. Rogach , and A. L. Bradley,
J. Phys. Chem. C \textbf{116}, 26529 (2012).


%




\bibitem{nitzan-cpl84} J. I. Gersten and A. Nitzan, Chem. Phys. Lett. \textbf{104}, 31 (1984).

\bibitem{nitzan-jcp85} X. M. Hua, J. I. Gersten, and A. Nitzan, J. Chem. Phys. \textbf{83}, 3650 (1985).

\bibitem{druger-jcp87} S. D. Druger, S. Arnold, and L. M. Folan, J. Chem. Phys. \textbf{87}, 2649 (1987).

\bibitem{dung-pra02} H. T. Dung, L. Kn\"{o}ll, and D.-G. Welsch, Phys. Rev. A \textbf{65}, 043813 (2002). 

\bibitem{stockman-njp08} M. Durach, A. Rusina, V. I. Klimov, and M. I. Stockman, New J. Phys. \textbf{10}, 105011 (2008).

\bibitem{pustovit-prb11} V. N. Pustovit and T. V. Shahbazyan, 
Phys. Rev. B \textbf{83}, 085427 (2011).

\bibitem{pustovit-prb13} V. N. Pustovit and T. V. Shahbazyan, 
Phys. Rev. B \textbf{88}, 245427 (2013).


\bibitem{schatz-jpcl17} L.-Y. Hsu, W. Ding, and G. C Schatz, 
J. Phys. Chem. Lett. \textbf{8}, 2357-2367 (2017).



\bibitem{moreno-nl10}  D. Martin-Cano, L. Martin-Moreno, F. J. Garcia-Vidal, and E. Moreno, Nano Lett. \textbf{10}, 3129 (2010).

\bibitem{velizhanin-prb12} K. A. Velizhanin and T. V. Shahbazyan, 
Phys. Rev. B \textbf{86}, 245432 (2012).

\bibitem{feldmann-nl05}E. Dulkeith, M. Ringler, T. A. Klar, J. Feldmann, A. M. Javier, and W. J. Parak, Nano Lett. \textbf{5}, 585 (2005).

\bibitem{novotny-prl06}P. Anger, P. Bharadwaj, and L. Novotny, Phys. Rev. Lett. \textbf{96}, 113002 (2006).

\bibitem{sandoghdar-prl06}S. K\"{u}hn, U. Hakanson, L. Rogobete, and V. Sandoghdar, Phys. Rev. Lett. \textbf{97}, 017402 (2006).

\bibitem{halas-nl07} F. Tam, G. P. Goodrich, B. R. Johnson, and N. J. Halas, Nano Lett. \textbf{7}, 496 (2007).

\bibitem{nitzan-jcp81}J. Gersten and A. Nitzan, J. Chem. Phys. \textbf{75}, 1139 (1981).

\bibitem{ruppin-jcp82}R. Ruppin, J. Chem. Phys. \textbf{76}, 1681 (1982).

\bibitem{silbey-review} R. R. Chance, A. Prock, and R. Silbey, 
Adv. Chem. Phys. \textbf{37}, 1 (1978)

\bibitem{huber-prb85} D. L. Huber, 
Phys. Rev. B \textbf{31}, 6070 (1985).

\bibitem{parkinson-prb07} P. Parkinson, E. Aharon, M. H. Chang, C. Dosche, G. L. Frey, A. K\"{o}hler, and L. M. Herz,
Phys. Rev. B \textbf{75}, 165206 (2007).

\bibitem{noginov-nanophot21} S. Rout, S. R. Koutsares, D. Courtwright, E. Mills, A. Shorter, S. Prayakarao, C. E. Bonner, and M. A. Noginov,
Nanophotonics \textbf{10}, 3659–3665 (2021).

\bibitem{noginov-josa19} S. Prayakarao, D. Miller, D. Courtwright, C. E. Bonner, and M. A. Noginov, 
J. Opt. Soc. Am. B \textbf{36}, 2312-2316 (2019). 

\bibitem{noginov-josa21} S. Koutsares, L. S. Petrosyan, S. Prayakarao, D .Courtwright, C. E. Bonner, T. V. Shahbazyan, and M. A. Noginov,
J. Opt. Soc. Am. B \textbf{38}, 88-94 (2021). 

\bibitem{noginov-nanomaterials20} 
S. Rout, Z. Qi, L. S. Petrosyan,T. V. Shahbazyan, M. M. Biener, C. E. Bonner, and M. A. Noginov,
Nanomaterials \textbf{10}, 2135 (2020).

\bibitem{silbey-cpl95} M. Cho and R. J. Silbey
Chem. Phys. Lett. \textbf{242}, 291-296 (1995).





%
%


\end{thebibliography}
\end{document}